# Differential Error Region of a Quantity Dependent on Full One-Port Network Analyser Measurements

N.I. Yannopoulou and P.E. Zimourtopoulos

An analytical method was developed to estimate errors in quantities depended on full one-port vector network analyser (VNA) measurements using differentials and a complex differential error region (DER) was defined. To evaluate the method, differences instead of differentials were placed over a DER which was then analysed and compared with another commonly used estimated error. Two real differential error intervals (DEIs) were defined by the greatest lower and least upper bounds of DER projections. To demonstrate the method, a typical device under test (DUT) was built and tested against frequency. Practically, a DER and its DEIs are solely based on manufacturer's data for standard loads and their uncertainties, measured values and their inaccuracies.

*Introduction*: In full one-port measurements with a VNA of real characteristic impedance $Z_0$, a DUT with impedance $Z$ has a reflection coefficient $\rho$ defined by

$$\rho = (Z - Z_0)/(Z + Z_0)$$

and related to its measured value m by the bilinear transformation

$$\rho = (m - D)/[M(m - D) + R]$$

in terms of errors D, M and R [1]. This transformation can be uniquely determined from given distinct $\rho_n$, n = 1, 2, 3 and respectively known $m_k$, k = n [2].

*Theory*: We considered $\rho_n$, $m_k$ as the elements of given ordered triples (A,B,C), (a,b,c), solved the resulting system and appropriately expressed its solution by

$$F = \sum cC(B - A)$$

$$D = \sum abC(A - B)/F$$

$$M = \sum c(B - A)/F$$

$$R = [\prod (A - B)(a - b)]/F^2$$

where $\sum$ and $\prod$ produce two more terms from the one shown, by rotation of the ordered triple elements. These errors were then considered as depended on the independent variables $\rho_n$, $m_k$. Therefore, their differentials were expressed in the same manner by

$$dD = [\prod (a-b) \sum (B - C)BCdA + \sum (b - c)^2(B - A)(C - A)BCda]/F^2$$

$$dM = [\sum (a - b)(c - a)(B - C)^2 dA - \prod (A - B) \sum (b - c)da]/F^2$$

$$dR = \{\sum [F + 2(a - b)B(A - C)][(B - C)^2 dA \prod (a - b) - (b - c)^2 da \prod (A - B)]\}/F^3$$

After that, the differential of $\rho$ was expressed by

$$d\rho = [-RdD - (m - D)^2 dM - (m - D)dR + Rdm]/[M(m - D) + R]^2$$

and was considered depended, through dD, dM and dR, on L = 7 independent variables and their independent differentials: $\rho_n$, n = 1, 2, 3 and $m_k$, k = n or k = 0 with $m_0$ = m.

The developed expressions were mechanically verified using a developed software program for symbolic computations.

*Application*: Manufacturer's data for standard loads used in full-one port VNA measurements are substituted in $\rho_n$, and for their uncertainties in $d\rho_n$. Since $Z_0$ is real, the domain of each $\rho_n$ is the closed unit circle [3]. For $|\rho_n|$ = 0 or 1, care must be exercised to restrict its differential value onto its domain. The VNA measurements have specified bounded ranges for their modulus and argument, so that the domain of each $m_k$ is a bounded circular annular with its centre at the origin O of the complex plane. Measurement data are substituted in $m_k$ and manufacturer's data for measurement inaccuracy in $dm_k$. Uncertainty and inaccuracy data outline domains for $d\rho_n$ and $dm_k$. If $z = |r|e^{j\varphi}$, stands for any of the independent variables and dz for its differential then the contribution of dz to d$\rho$ is a summation term of the form Wdz, with W = $|U|e^{jV}$, so that

$$Wdz = |U|e^{j(V + \varphi)}d|r| + |U|e^{j(V + \varphi + \pi/2)}|r|d\varphi$$

where W is in fact a known value of the respective partial derivative and $d|r|$, $d\varphi$ are the independent real differentials of the complex dz in polar form. Each expression Wdz outlines a contour for a partial DER around O. If z ≠ 0, the partial DER is a parallelogram with perpendicular sides $d|r|$ and $|r|d\varphi$, stretched or contracted by $|U|$ and rotated by (V + $\varphi$) around O. If z = $\rho_n$ = 0, the partial DER is a circle with radius $|U|d|r|$. Accordingly, a DER is the sum of either L parallelograms or (L - 1) parallelograms and 1 circle. DER is then a convex set with contour either a polygonal line with 4L vertices at most, or a piecewise curve composed of 4(L - 1) line segments and 4(L - 1) circular arcs at most. The greatest lower and least upper differential error bounds are the end-points of DEIs for the real and imaginary parts of d$\rho$ and result from the projections of DER for $\rho$ on the coordinate axes.

These conclusions can be generalized for any other quantity directly or indirectly depended on all, some or just one of the above independent variables and their differentials. Thus, the quantity has an L-term DER, where $7 \geq L \geq 1$. For example, the impedance Z of a DUT has the 7-term DER:

$$dZ = 2Z_0 d\rho/(1 - \rho)^2$$

*Results*: All of the following data are specified by manufacturers of the parts for our measurement system. This system operates from 1 to 1300 MHz with 100 Hz PLL stability and consists of a type-N $Z_0 = 50$ $\Omega$ network analyser, a number of support instruments and a set of standard loads. The standards are: a short circuit A, a matching load B with reflection coefficient 0.029 and an open circuit C with reflection coefficient 0.99 and phase accuracy $\pm 2°$. In the absence of manufacturer's data for A we considered its uncertainty equal to that of C. So, the following values were substituted in the developed expressions: A = −1, $0 \leq d|A| \leq 0.01$, $−180° \leq d\varphi_A \leq −178°$ or $178° \leq d\varphi_A \leq 180°$, B = 0, $|dB| = 0.029$, C = 1, $−0.01 \leq d|C| \leq 0$, $−2° \leq d\varphi_C \leq +2°$. The annular domain for $m_k$ of VNA is specified from 0 to -70 db in modulus and $\pm 180$ degrees in argument. Measurements $m_k$ result with a decimal floating point precision of 4 digits, for both modulus and argument. We consider the modulus and argument of $dm_k$ equal to $\pm 1/2$ of the unit in the last place of the corresponding mantissa in modulus and argument of $m_k$. Consequently, our system produces a

DER, either for $\rho$ or Z, as a sum of (L -1) = 6 parallelograms and 1 circle, with a contour of (4L + 4L) = 48 vertices at most.

A suite of developed software applications: (i) controls the system and collects the data in terms of frequency using the IEEE-488 protocol, (ii) processes the collected data and computes the vertices of DER and the end-points of its DEIs (iii) sketches pictures of DER for $\rho$ and its counterpart Z in terms of the frequency steps and make a film using them as frames.

A typical resistor with a nominal DC impedance of 50 $\Omega$ $\pm$20% tolerance was soldered on a type-N base connector and enclosed in an aluminium box to serve as a simple DUT for testing its Z from 2 to 1289 MHz in 13 MHz steps. The centre frequency $f_C$ = 639 MHz was chosen to reveal the details of the proposed method in Fig. 1, where the contour of a typical DER for Z is outlined with small circles as its vertices. This contour surrounds that of the 4-terms DER due to inaccuracy of measurements (1) and that of 3-terms DER for the uncertainty of loads (2). A properly circumscribed rectangle of DER shows graphically how the DEIs for R and X result. The commonly used error from the matching load only is shown as a dotted circle. This is in fact a 1-term DER which is surrounded from the contour of the DER by a factor of about 125% to 185% in all directions. Finally, in the same figure, $2^{7 \times 2}$ differences $\Delta Z$ resulting from the same $d\rho_n$ and $dm_k$, dense enough to appear as stripes, were placed over DER to compare them with differential dZ values. Notably, almost all of $\Delta Z$ values are belong to DER while the computation time for these $\Delta Z$ exceeds that for DER by more than one order of magnitude. To demonstrate the method, a set of selected DER frames for $\rho$ and Z are shown in Fig. 2, as beads on space curved filaments against frequency. Finally, the computed DEIs for R and X are shown in Fig. 3 against frequency.

The proposed method may be efficiently used in the same way, to successfully estimate errors in any quantity depended on full one-port vector network analyser measurements.

Authors' affiliations:

N.I. Yannopoulou and P.E. Zimourtopoulos (Antennas Research Group, Microwaves Laboratory, Section of Telecommunication and Space Science, Department of Electrical and Computer Engineering, Democritus University of Thrace, V. Sofias 12, Xanthi, 671 00, Greece)

E-mail: yin@antennas.gr

Figure captions:

Fig. 1 A typical differential error region DER for the impedance Z

Fig. 2 DER for the reflection coefficient $\rho$ and for its associated impedance Z against frequency

Fig. 3 Greatest lower and lest upper differential error bounds for resistance R and reactance X against frequency

Figure 1

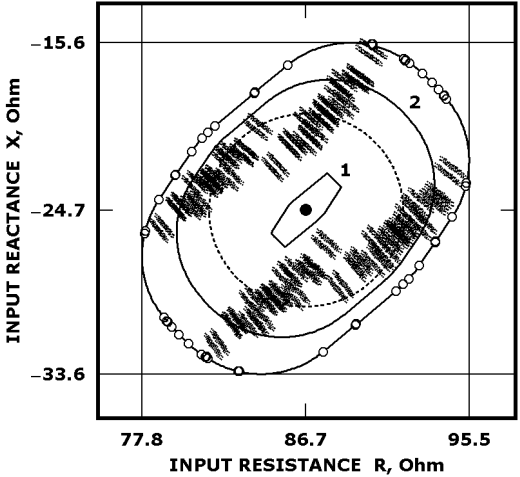

Figure 2

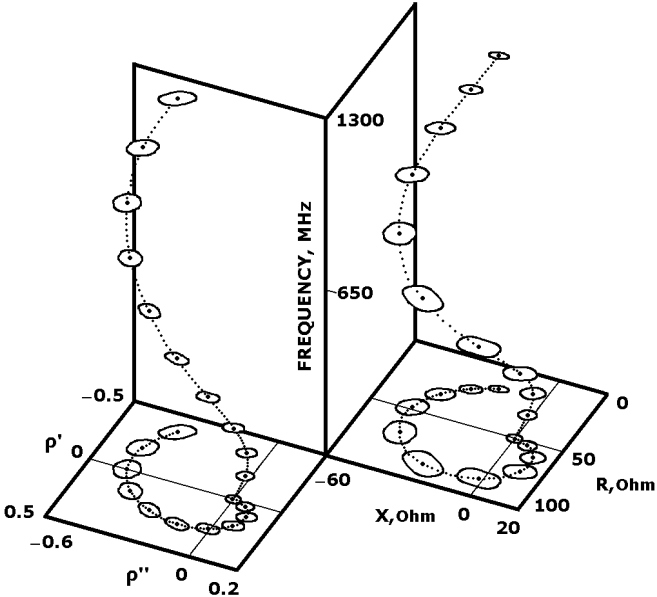

Figure 3

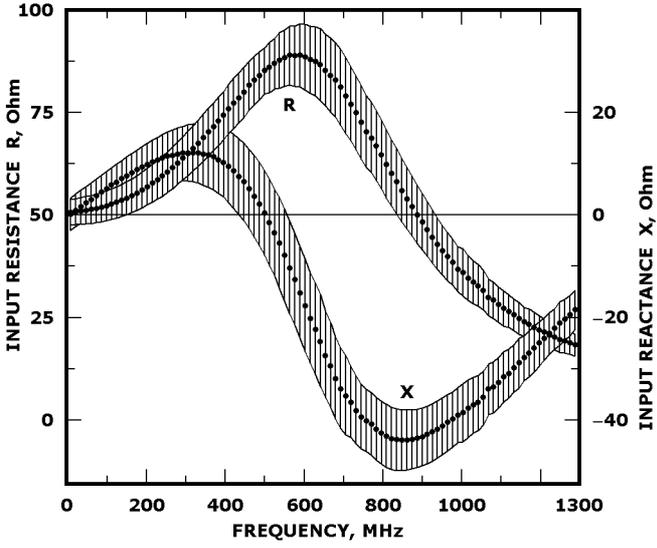